\documentclass[12pt]{article}

\input epsf
\def\beq{\begin{eqnarray}}
\def\eeq{\end{eqnarray}}

\def\mpl{M_{\rm Pl}}

\def\l{{\cal L}}

\def\lsim{\mathrel{\rlap{\lower3pt\hbox{\hskip0pt$\sim$}}
     \raise1pt\hbox{$<$}}}         %less than or approx. symbol
\def\gsim{\mathrel{\rlap{\lower4pt\hbox{\hskip1pt$\sim$}}
     \raise1pt\hbox{$>$}}}         %greater than or approx. symbol

\begin{document}

%\begin{titlepage}

\begin{flushright}
{ NYU-TH-04/12/13}
\end{flushright}
\vskip 0.9cm

\centerline{\Large \bf Lorentz-violating massive gauge and 
gravitational fields}

\vskip 0.7cm
\centerline{\large Gregory Gabadadze and Luca Grisa}
\vskip 0.3cm
\centerline{\em Center for Cosmology and Particle Physics}
\centerline{\em Department of Physics, New York University, New York, 
NY, 10003, USA}

\vskip 1.9cm

\begin{abstract}

We study nonlinear dynamics in models of Lorentz-violating massive gravity.
The Boulware-Deser instability restricts severely the class of acceptable 
theories. We identify a model that is stable. It exhibits
the following bizarre but interesting property: there are only two massive 
propagating degrees of  freedom in the spectrum, and  yet long-range 
instantaneous  interactions are present in the theory. We 
discuss this property on 
a simpler example of a photon with a Lorentz-violating mass term 
where the issues of (a)causality are easier to understand. 
Depending on the values of the mass parameter these models can either 
be excluded, or become phenomenologically interesting. 
We discuss a similar example with more degrees of freedom, 
as well as a model without the long-range instantaneous interactions.

\end{abstract}

\vspace{3cm}

%\end{titlepage}

\newpage

\section{Introduction and summary}

Models of a massive/metastable graviton could shed  
some light on the cosmological constant problem
(see, e.g., \cite {GG}). Given the ultraviolet problems of 
gravity, provisionally, one would like to find a model that 
could be regarded as a classically and quantum mechanically 
consistent {low-energy} effective field theory (for a 
description of gravity as an effective theory see Ref. \cite {Don}).
In searching for such a theory, that could also preserve Lorentz 
invariance in four-dimensions, one typically encounters  
the following three major problems:

{\it Problem 1. Linear discontinuity.}  Known as the van 
Dam-Veltman-Zakharov (vDVZ) discontinuity \cite {Veltman,Zakharov}. 
To understand this problem let us ignore for the time being  
all possible nonlinear self-interactions of a spin-2 field.
The Lagrangian of this theory without ghosts and tachyons was 
uniquely determined by Fierz and Pauli (FP) \cite {PF}. Irrespective of the 
details of the Lagrangian, Lorentz invariance dictates
that a massive spin-2 state has to have five physical polarizations,
as opposed to a massless graviton with only two polarizations.
This is what gives rise to the vDVZ discontinuity: one out of the extra three
degrees of freedom couples to the trace of the stress-tensor, and, 
no matter how small the graviton mass, gives rise to 
experimentally unacceptable predictions either for the light 
bending by the Sun or for Newtonian interactions. 
Thus, the FP Lagrangian  gives a {\it theoretically} consistent model of 
a massive spin-2 state without self-interactions, however, if this spin-2 
state is declared  to be a graviton, the model is in contradiction 
with observations. 

{\it Problem 2. Strong coupling.} The above arguments 
were based on the linearized approximation.
In a theory of gravity we should take into account nonlinearities. 
It was  first observed by Vainshtein \cite {AV} that the arguments 
leading to the vDVZ discontinuity fail once  the nonlinear interactions are 
taken into account. This is because linearized approximation breaks down, 
and, to make predictions within 
the Solar system one should either solve the nonlinear equations exactly,
or come up with an alternative viable approximation.
The latter two approaches can restore the consistency of 
the predictions of classical massive gravity with observations 
\cite {AV,DDGV}. The breakdown of the linearized approximation 
takes place because of the nonlinear but classical self-interactions of the 
extra polarizations \cite {DDGV}. This suggests a problem  for quantum theory
where the same nonlinear interactions appear in the loop diagrams, 
leading eventually to a very low ultraviolet cutoff \cite {AGS} 
for a quantum graviton scattering on an empty background. 
Moreover, quantum loops are expected to generate  higher dimensional 
operators that are suppressed only by this observationally 
unacceptable low cutoff \cite {AGS}. Note that in a theory of gravity 
a cutoff has no  universal meaning. For instance, for graviton 
scattering in a background of classical sources the physical cutoff 
would depend on local curvature invariants. This in principle could 
be used to try to overcome the strong coupling problem (along the lines of 
\cite{Dvali,Nicolis}) if the model were consistent otherwise. 
This brings us to the issue of a nonlinear instability of the FP model.

{\it Problem 3. Nonlinear instability.} Also known as the Boulware-Deser (BD) 
instability. From our standpoint, clarified below, 
this is the most severe problem. It emerges already at the 
{\it classical} level. To 
quickly sketch the essence of the BD instability
one can look at a scalar field model with the Hamiltonian
$H=(\partial_t\phi)^2+(\nabla \phi)^2 + m^2\phi^2 + \sigma 
((\partial_t\phi)^2+(\nabla \phi)^2) +m^3\sigma^2 \phi/2$. 
In the quadratic approximation this describes a single free scalar 
field $\phi$. However, because of the nonlinear 
(cubic in this case) terms  the instabilities set in. 
One can simply integrate out
the $\sigma$ field and obtain that the Hamiltonian contains the 
term $-((\partial_t\phi)^2+(\nabla \phi)^2)^2/2m^3\phi$, which  
is sign-indefinite and unbounded below.  This shows the essence of the BD 
instability in an oversimplified model (for complete treatment,
see section 3). In this toy model the problem can be 
cured by adding new terms into the Hamiltonian. One might hope 
for a similar  remedy  in the FP theory.  Indeed, a  
nonlinear completion of the FP gravity 
is not unique and, in any event, one should expect quantum 
loops to generate all sorts of new nonlinear terms.  
There are examples of nonlinear systems where certain 
classical instabilities are removed by quantum--loop--generated 
terms\footnote{For instance, some models   
exhibit chaotic behavior at the classical level while 
the  chaos is eliminated by quantum corrections.}. 
However, to the best of our knowledge, 
there is no evidence for this to be happening 
in the FP model. In particular: (a) The worst part of the BD instability 
exists because of the {\it nonlinear} interactions  between 
the tensor sector of  the conventional Einstein-Hilbert Lagrangian
and the extra polarizations of a massive graviton that  
are only present in the mass term (i.e., the Nambu-Goldstone (NG) 
sector of massive gravity). 
(b) It was shown in Refs. \cite {BD,GGruzinov} that 
the BD type instabilities persist  for {\it an arbitrary} polynomial 
in {\it fields} completion of the FP gravity. Note that these terms include 
all sorts of derivatives of the Nambu-Goldstone boson. 
(c) One might hope that these instabilities go away once the 
terms that contain derivatives of fields are included. However, this 
would require an infinite number of fine tunings
for which no symmetry principle is known in the four-dimensional 
context \footnote{One can  also try to modify the linear part of the FP 
theory by introducing heavy states  at the UV cutoff of the theory.  
Explicit calculations, similar to those of section 3, show 
that the rapid BD instability persists in this case too.
We thank Nima Arkani-Hamed for suggesting to study this question.}.
The unstable solutions found in Ref. \cite {GG} were of a cosmological
type. It would be interesting to conduct similar studies
for spatially inhomogeneous localized sources.

Summarizing, the BD instability 
is the  most severe problem:  unlike Problems 1 and 2, it questions the 
very consistency of a theoretical model itself. 
So far  no concrete cure was proposed in the context of 
{a Lorentz-invariant local field theory} with a finite number of 
degrees of freedom and without an infinite number of fine tunings.
Surely, a consistent theory of nonlinearly interacting massive spin-2 
should exist, it just has not been formulated yet.
Therefore,  until shown otherwise, we will assume that the models 
with the DB instability  should be avoided.

In an ideal case, one would like to have a model 
in which all the above three problems are absent.
However, if one should compromise between Problems 1,2 and 3, 
as it will be the case in one particular  examples below, 
our approach will be to worry first of all about  the Problem 3.
This is because Problems 1 and 2, although unpleasant from the
point of view of practical calculations, can be taken care of
consistently. For instance, in the DGP model  
\cite {DGP} these problems are solved 
at the classical \cite {DDGV} as well as at 
the quantum level \cite {Dvali,Nicolis}\footnote{We also 
note that there exist models that have no Problems 2 or 3 
\cite {GS,Romb}, however they  exhibit 
the vDVZ discontinuity. A version of \cite {DGP} discussed in \cite {GGweak} 
could potentially evade all the problems in the weak 
coupling regime, however some nonlinear issues should 
still be understood in that approach (see the 
discussions section in \cite{GGweak}).}.

Recently, a new approach to massive gravity was initiated in 
Ref. \cite {Ghost}: the idea is to give up Lorentz-invariance which 
could be spontaneously broken when graviton acquires its mass \cite {Ghost}. 
Subsequently, in Ref. \cite {Rubakov} a general parametrization of the 
Lorentz-violating (LV) graviton mass term was proposed and the 
models evading the  Problems 1 and 2 were identified. 
More general studies of the proposal 
of \cite {Rubakov} were performed in Ref. \cite {Dubovsky}.
The discussions in  Ref. \cite{Rubakov} were restricted 
to the linearized theory. 
The purpose of the present work is to study a complete 
nonlinear dynamics in the models with  the LV mass terms,
and in particular address the Problem 3. 
We will find that many of the LV mass  models suffer
from the BD instability. However, there are at least 
two classes of models that can evade the problem. 
The first class exhibit surprising properties: 
even though there are only massive propagating degrees of freedom, 
the models exhibits a long-range 
instantaneous interactions. This could be phenomenologically deadly or 
interesting depending on the value of the graviton mass. 
It is very  likely that the properties of some of these  
models won't be affected  by radiative corrections since they  
are protected by certain symmetries. (A different model
but with somewhat similar properties was discussed in 
\cite {Dubovsky} in the linearized approximation, see also 
\cite {Dubovsky1}). The second class of the BD stable models 
contains all the massive degrees of freedom and no 
long range interactions. However, the issue of radiative corrections for 
these models remain open, without an infinite number of fine tunings,
these models are likely to  exhibit the instabilities at the quantum level.

The work is organized as follows. In section 2, as an instructive 
example, we discuss a Lorentz-violating theory of a massive photon. 
This model contains only massive propagating degrees of freedom
(two massive polarizations of an electromagnetic wave), and
nevertheless, there are long-range instantaneous interactions
in the theory. We briefly discuss whether this type of 
models can be consistent with observations. In section 3 we overview
the BD instability in the FP theory and show how it 
also appears in the LV models. In section 4, first  we discuss a model
that has no BD instability and propagates only two massive
degrees of freedom (transverse polarizations of gravitational 
waves). We study the long-range interactions in this model,
showing that there is no vDVZ discontinuity. Finally we discuss two
other models, one with 5 massive degrees of freedom and 
the long-range-interactions, and another one with 
six degrees of freedom where all the interactions 
are screened.

\section{Warming up with photons}

As a toy but very interesting example we consider QED with 
Lorentz-violating mass term for a photon\footnote{Michele Papucci 
and Matthew Schwartz also studied this model for a photon. 
We thank Matthew Schwartz for communications on this.}
(for convenience we call it QED')
\beq
\l=-\frac{1}{4}F_{\mu\nu}F^{\mu\nu}-\frac{1}{2}m^2A_j A^j \,-\,A_\mu J^\mu\,,
\label{qed1}
\eeq
where $\mu,\nu=0,1,2,3; ~i,j=1,2,3$; $J^\mu$ is a conserved current  
$\partial^\mu J_\mu=0$, and  the mass term 
breaks explicitly the Lorentz group down to the group of three 
dimensional spatial rotations $SO(3)$ (our choice of the Lorentzian 
signature is $[-+++]$.). One could think of this model
as arising from a Lorentz-invariant theory in which
certain fields acquire Lorentz-violating VEV's (see, e.g., \cite {Ghost},
\cite {Grip}). These VEV's set a preferred frame in the Universe 
in which the model (\ref {qed1}) is defined.  

The Lagrangian (\ref {qed1}) is invariant 
under spatially independent
gauge transformations of the fields $\delta A_0 =\partial_0 \lambda(t),~~ 
\delta A_j =0$. Because of this and conservation of 
$J_\mu$ no new terms are generated by quantum loops in 
(\ref {qed1}).  The equations of motion of  the  model are
\beq
\partial^\nu F_{\nu\mu}-m^2\delta_{\mu}^iA_i=J_\mu \,.
\label{eq1}
\eeq
There are two key properties that follow from (\ref {eq1}).
First, the zeros component of this equation implies that 
the Gauss's law is identical to that of QED:
\beq
\partial^j E_j =-J_0\,,
\label{gauss}
\eeq
where $E_j=F_{0j}$ is an electric field and $J_0$ is a charge density.
Hence, in QED', like in QED, the electric field is {\it not screened}! 
Second, taking the partial derivative of both sides of (\ref {eq1})  we 
find a  Lorentz-violating analog of the Proca condition
\beq
\partial^iA_i= 0\,.
\label{proca}
\eeq
As a result, there remain  only two dynamical
degrees of freedom in the theory: $A_0$ is not dynamical
and one of the $A_j$'s can be expressed though the other two using 
(\ref {proca}). Both propagating degrees of freedom are massive. 

However, this is not all. We note that (\ref {proca}) 
coincides with the {\it Coulomb} gauge fixing condition of QED. 
Because of this, a free photon propagator of (\ref {qed1}) 
\beq
D_{00}(k)\,=\,{1\over {\vec k}^2},~~~D_{i j}(k)\,=\,\left (\delta_{ij} 
- {k_ik_j\over {\vec k}^2} \right ) {1\over -k_0^2 +{\vec k}^2+m^2  
-i\epsilon}\,,
\label{prop}
\eeq
resembles a causal Coulomb gauge QED propagator.
The physics of QED' (\ref {qed1}) is rather different, however. 
Like in QED, there is an  instantaneous Coulomb potential 
(the zero-zero component of the propagator). 
This component has no imaginary part, hence, there are no 
physical degrees of freedom mediating the instantaneous potential.
The spatial components, on the other hand, have an imaginary part. This  
corresponds to the two physical degrees of freedom. 
Unlike in QED, in the present case both 
of these degrees of freedom are massive. This has a dramatic consequence: 
as we will see shortly, the instantaneous potential  
is not canceled in physical observables for time 
dependent sources.  The remaining instantaneous field is small for 
transverse sources with a typical momentum/frequency scales 
$ \gg m $, but  can become essential  for scales  $\lsim  m $
generating {\it action-at-a-distance}. To examine this question in 
detail we follow closely  massless QED 
in {\it Coulomb gauge}. In the latter  
case one postulates  (\ref {proca}) as a gauge condition. 
As a result of this
\beq
A_0(r,t)\,=\,{1\over 4\pi}\,\int d^3 r' { J_0(r',t) \over |r-r'|}\,,
\label{a0}
\eeq
is an instantaneous potential. The expression (\ref {a0}) 
is identical in QED and QED'.
On the other hand, the equation  for the vector potential 
differ in the two models. The spatial part of (\ref {eq1}) reads: 
\beq
(\partial^2 -m^2) A_j =J_j - \partial_j\partial_0 A_0\,\equiv \,J_j^{tr}\,.
\label{aj}
\eeq
The mass term on the l.h.s. is present only in QED' but not in QED. 
Both in  QED and QED' the vector potential $A_j$ has 
an instantaneous parts. In QED this part exactly 
cancels  (\ref {a0}) in physical observables such as 
the electric field $E_j=-\partial_jA_0 + \partial_0A_j$.
However, this cancellation is not exact in QED'.
To see this we write:
\beq
 A_j(r,t)= \int d^3r'dt' D_R(r-r';t-t') J_j^{tr}(r',t')\,,
\label{ajex}
\eeq
where the retarded Green's function
\beq
D_R(r;t')\equiv  D^{QED}_R\,+\,D^{m}_R = {\theta(t)\over 2\pi}\,
\delta(t^2-{\vec r}^2) -
{\theta(t-|r|)\over 4\pi}\, { m \,{\cal J}_1(m \sqrt{t^2-{\vec r}^2})\,
\over \sqrt{t^2-{\vec r}^2}} \,,
\label{ret}
\eeq
is expressed through the step function $\theta$, Dirac 
delta function $\delta$,  
and Bessel function ${\cal J}$ \cite{BS}. Note that the mass dependence in the
Green's functions is additive. We used this to denote the massless 
function by $D^{QED}_R$ and the addition due to the mass term by $D^m_R$. 
Using (\ref {ret}) we can  write
\beq
A_j(r,t)=A^{QED}_j(r,t)+ \int d^3r'dt' D^m_R(r-r';t-t') 
J_j^{tr}(r',t')\,,
\label{aj1}
\eeq
where $A^{QED}_j(r,t)$ is a vector potential of massless QED in Coulomb gauge.
The latter, as we mentioned before,  cancels the instantaneous 
scalar potential (\ref {a0}) to produce  the {\it retarded} electric 
field $E_j^{QED}$. Therefore,
\beq
E_j (r,t)= E_j^{QED}(r,t)+\partial_0 \int d^3r'dt' D^m_R(r-r';t-t')\,
J_j^{tr}(r',t')\,,
\label{ej}
\eeq
and the instantaneous part is now contained only in 
the last term of this expression. The latter  
can be calculated as follows:
\beq
{i\over (2\pi)^4} \,\int d^3k\, e^{i{\vec k}{\vec r}}\, {\rm Re}
\left (e^{i \,\sqrt{k^2+m^2}\,t}\, {\tilde J}^{tr}_j({\vec k},\sqrt{k^2+m^2})-
e^{i \,|k|\,t}\, {\tilde J}^{tr}_j({\vec k},|k|) \right )\,,
\label{deltaE}
\eeq
where 
\beq
{\tilde J}^{tr}_j({\vec k},\omega)\equiv \int d^3rdt \, 
e^{i{\vec k}{\vec r}-i\omega t}
J_j^{tr}({\vec r},t)\,,
\label{ft}
\eeq
is the Fourier transform of the transverse current\footnote{
For simplicity (\ref {deltaE}) is written for a source such that 
${\tilde J}(k,-\omega)= {\tilde J}(k,\omega)$. However, a 
general expression can readily be obtained.}.

In general, the expression (\ref {deltaE}) is nonzero 
even for $t=0$.  It appears that 
an information from an event  that took place at $t=0, r=0$ can 
{\it instantaneously} be transmitted  
to a point that is far away from this location.
This gives rise to the action-at-a-distance.
The question how important this instantaneous 
interaction is depends on properties of  sources.
For a  transverse source of a typical momentum $k_0$ 
and typical frequency $\omega_0$ the effect is negligible 
as long as $k_0\gg m$. For $k_0 \ll m$ the effects 
can be appreciable when $\omega_0\sim m$ or $\omega_0\sim k_0$.
In this case the instantaneous electric  field would decay 
with distance as $\sim 1/r$. If the source has no typical frequency,
juts a typical momentum, then the instantaneous interactions 
will  be important for $k_0\ll m$, and vice versa, 
for a source of a typical  frequency $\omega_0$ and no typical 
momentum the dangerous interactions will be present for 
$\omega_0\lsim m$. In practice, to produce a low-momentum/frequency 
signal that could trigger the instantaneous interaction, will 
itself take certain characteristic time. It 
would be  interesting to study the phenomenology of these
interactions for realistic sources to put bounds on $m$ 
\cite  {GLuca}. Note also that magnetic field has no instantaneous 
parts in QED': 
${\vec B} = {\rm curl} {\vec A}$ and the curl eliminates the 
instantaneous part of ${\vec A}$. 

The presence of the long-range interactions can also 
be understood from the Hamiltonian 
formulation of (\ref {qed1}) where  $A_0$  acts as a 
Lagrange multiplier:
\beq
{\cal H}= {1\over 2} P_j^2 \,+\,{1\over 2} (\epsilon_{ijk}\partial_jA_k)^2
+{1\over 2} m^2\,A_j^2 + A_0 \left ( \partial_j P_j - J_0  \right ) +A_jJ_j\,.
\label{ham0}
\eeq
Here $P_j=F_{j0}$ is the canonical momentum.
Variation w.r.t. $A_0$  gives rise to the Gauss's law, 
$\partial^j P_j = J_0$, which is identical to the Gauss's law
of massless QED.  This guaranties that the 
theory possesses the long-range interactions in spite of the 
fact that the two dynamical propagating degrees 
of freedom are massive\footnote{The condition $\partial^j A_j =0$
is obtained by taking a derivative of one of the Hamiltonian 
equations of motion.}. 

To summarize, there are two massive propagating degrees of freedom, however, 
there still exists a long-range instantaneous 
interaction in the theory. Depending on the value of $m$ this 
can exclude a given 
model, or be potentially interesting for phenomenological 
applications.

One can of course modify the model (\ref {qed1}) by adding a 
``mass term'' $\alpha\, m^2 A_0^2$ with some nonzero positive 
coefficient $\alpha$. In this case the long-range 
interactions are removed from the theory ($\alpha=1$ corresponding to 
a massive Lorentz invariant photon). However, it is worth pointing out that 
$\alpha=0$ is an enhanced gauge symmetry point and 
unlike the $\alpha\neq 0$ cases should be stable under radiative 
corrections\footnote{The $\alpha=1$ case is also stable because of the 
restored Lorentz invariance and conservation of the current.}.

\section{Nonlinear instabilities in massive gravity}

In this section we summarize how nonlinear instabilities appear
in a Lorentz-invariant theory of massive gravity (the FP gravity) 
\cite {BD}, and show that the similar instabilities exist in 
many of the LV mass models once the nonlinear interactions 
are taken into account.   We will also find the conditions 
under which these instabilities can be removed in a Lorentz-violating
massive theory. In the latter case, however, one typically ends 
up with long-range interactions, similar to those studied in 
the previous section. In this respect the theory is half-massive:
all its degrees of freedom are massive nevertheless there are long-range 
interactions.  One atypical example that evades the instabilities 
and long-range instantaneous interactions will be given at the end of the 
next section.

We first review briefly the Hamiltonian construction of \cite {BD}
to identify the terms that are responsible for the 
instabilities. Then, we will remove these terms in a 
Lorentz-violating theory. Let us start by a brief reminder  
of the ADM formalism \cite{ADM}. This would be a 
natural formalism for nonlinear formulation of the 
LV mass gravity. Consider a foliation of space-time by hyper-surfaces 
$\Sigma_t$ parametrized by a time variable $t$. 
The four-dimensional metric is replaced by the following 
three-dimensional variables:
\beq
\gamma_{ij}\equiv g_{ij};\quad N\equiv(-\phantom{.}^{(4)}g^{00})^{-1/2};\quad
N_i\equiv\phantom{.}^{(4)}g_{0i}\,.
\eeq
In terms of these variables the invariant interval takes the form:
\beq
ds^2=-(N^2 -N_jN^j)dt^2+2N_jdx^j\,dt\,+\gamma_{ij}dx^i\,dx^j\,,
\label{int}
\eeq
where all the spatial indices are contracted by means of  
the three-dimensional metric $\gamma_{ij}$. 
$N$ is called the lapse function and $N_i$ is the shift 
function. With these definitions
\beq
\sqrt{-\phantom{.}^{(4)}g}=N\sqrt{\gamma}\,,~~~~~
\phantom{.}^{(4)}R=\phantom{.}^{(3)}R+K_{ij}K^{ij}-K^2\,,
\eeq
where $K_{ij}$ is the extrinsic curvature of $\Sigma_t$
\beq
K_{ij}=\frac{1}{2}N^{-1}
\left[\dot\gamma_{ij}-\nabla_{i}N_{j} -\nabla_{j}N_{i}\right]\,,
\eeq
and $K$ is its trace. The extrinsic curvature is related to the  
canonical momentum 
\beq
\pi^{ij}\equiv\frac{\delta\l}{\delta\dot\gamma_{ij}}=\sqrt{\gamma}
(K^{ij}-K\gamma^{ij})\,.
\eeq
The Hamiltonian of the Einstein gravity in terms of the above variables reads
\beq
\mathcal{H}_{EH}=\pi^{ij}\dot\gamma_{ij}-\l=\sqrt\gamma
\left[NR^0+N_iR^i\right]\,,
\label{ham}
\eeq
where
\beq
R^0 \equiv -\phantom{.}^{(3)}R+\gamma^{-1}(\pi_{ij}\pi^{ij}-\frac{1}{2}\pi^2)\,
,~~~R^i \equiv -2\nabla_j(\gamma^{-1/2}\pi^{ij})\,.
\label{R0}
\eeq
$N$ and $N_i$ appear linearly in the Hamiltonian (\ref {ham}).
Hence they are Lagrange multipliers variation w.r.t. which  
gives the constraints $R^0=0$ and $R^i=0$;   
$\mathcal{H}_{EH}=0$ on the surface of the 
constraints.

Let us now turn to massive FP gravity for which the mass terms is written as
\beq
-\frac{1}{2}m^2(h_{\mu\nu}^2-{h_\mu^\mu}^2)=-\frac{1}{2}
m^2\left[h_{ij}^2-h^2-2N_i^2+2h(1-N^2-N^2_j)\right]\,,
\label{PF}
\eeq
where the second equality is obtained by expressing   
$h_{\mu\nu}=g_{\mu\nu}-\eta_{\mu\nu}$ in terms of
$\gamma_{ij}$, $N$ and $N_i$ (note that $h_{ij}=
\gamma_{ij}-\eta_{ij}$, and $h\equiv \gamma^{ij}h_{ij}$).
The key role is played by the terms in (\ref {PF}) which are quadratic in 
$N$ and $N_j$. Because of these terms $N$ and $N_j$  ceases 
to be  Lagrange multipliers in the massive theory. 
Variations w.r.t. $N$ and $N_j$ lead to the following equations:
\beq
\sqrt{\gamma}\,R^0 = 2m^2hN\,,~~~\sqrt{\gamma}
R^i = 2m^2(\eta^{ij}-h\gamma^{ij})N_j\,.
\label{const1}
\eeq
These are  not constraint equations any more but serve to 
determine $N$ and $N_j$. Substituting these solutions into 
the full Hamiltonian we obtain:
\beq
\mathcal{H}=\frac{1 }{4m^2}\left \{ { (\sqrt {\gamma} R^0)^2 \over h }+
	\gamma \,R^i(\eta^{ij}-h\gamma^{ij})^{-1}R^j \right \}+ 
\frac{1}{2}m^2(h_{ij}^2-h^2+2h)\,.
\label{hambd}
\eeq
This is a Hamiltonian of an ill defined theory. The first term 
on the r.h.s. is unbounded below and singular in $m$ and $h$.
For instance, consider $ \sqrt {\gamma}R^0$ fixed and $R^i=0$; 
when $h\rightarrow0^-$ 
the term in the Hamiltonian density $\mathcal{H}\sim (\sqrt {\gamma}
{R^0})^2/(m^2h)$
is not bounded below. This demonstrates the presence of 
a ghost-like instability in the theory.
This instability can manifest in many ways even at  the {\it classical} 
level, and the time scale of the instability can be very  short 
\cite {GGruzinov}. Such a theory is hard to make sense of.

The BD problem is associated with the terms 
that in the linearized theory looks as $h_{00}h^j_j$. 
The mass term of the model analyzed in Ref. \cite {Rubakov} is
\beq
L_m={\mpl^2\over 2} \left (m_0^2 h_{00}^2 +2 m_1^2 h_{0j}^2 -m_2^2 h_{ij}^2 +
m_3^2 h^2 -2 m_4^2h_{00}h \right )\,,
\label{massR}
\eeq 
with $m_0=0$ and all the other mass parameters being nonzero
with a certain hierarchy between them \cite {Rubakov}. A 
straightforward non-linear completion of the mass term (\ref {massR}) 
gives rise to  the BD instability. This is because
$h_{00} = 1 -N^2 +N_j^2$ and $N$ ceases to be a Lagrange multiplier,
as in the FP gravity. While in the FP model this instability cannot be removed
in a Lorentz-invariant fashion, in the present context  
Lorentz-invariance is broken anyway, and nothing prevents us to eliminate 
the dangerous term by judicially  choosing  $m_4$ to be zero.  
This choice is a point of enhanced gauge symmetry  and should be stable 
under loop corrections. However, the physics of the model with 
$m_0=m_4=0$ is dramatically different -- there appear long-range interactions. 
This will be discussed in detail in the next section.

\section{Stable models} 

{\it 1. Half-massive gravity} \\
We consider a simple Lorentz-violating generalization of the FP mass term: 
\beq
\Delta \l_1\,=\,-\frac{1}{2}\,m^2\,\sqrt{\gamma}N(h_{ij}^2-ah^2)\,,
\label{halfmass}
\eeq
where $a$ is a constant. As before, all the indices  are contracted by 
$\gamma_{ij}$. We think of this theory as being obtained
from a Lorentz-invariant model  through spontaneous generation 
of a preferred frame, similar in spirit to \cite {Ghost}.

The total Hamiltonian in this case takes the form:
\beq
\mathcal{H}=
\sqrt{\gamma}\left[
N\left (R^0+\frac{1}{2} m^2(h_{ij}^2-ah^2)\right)\,+\,N_i R^i  \right]\,.
\label{ham05}
\eeq
The constraint equations that follow are:
\beq
\label{N}R^0 = -\frac{1}{2}m^2(h_{ij}^2-ah^2)\,,~~~ \label{Ni}R^i= 0\,.  
\label{cons2}
\eeq
The Hamiltonian (\ref {ham05}) on the surface of constraints (\ref {cons2})
is zero, just like in the Einstein theory,  hence, the BD instability is gone.
The model (\ref {halfmass}) is invariant in the linearized 
approximation under 
coordinate-independent gauge transformations: $\delta h_{\mu\nu}=
\partial_\mu \zeta _\nu(t) + \partial_\nu \zeta _\mu(t)$, as well as
under the transformations with a gauge function  
$\xi_\mu = (\xi_0(t,{\vec x}), \xi_j=0)$ (at the nonlinear level there 
is a symmetry w.r.t. the spatially independent transformations). 
Due to this, we expect that the properties of this model will stay 
stable under  quantum loop corrections.

Let us now couple this theory to matter. The $\{00\}$ component of 
the equation of motion  takes the form (we use the units $\mpl=1$):
\beq
R^0\,+\,\frac{1}{2}m^2(h_{ij}^2-ah^2)\,=\,T_{00}\,.
\label{mass00}
\eeq
The $\{0j\}$ equations are identical to those of the Einstein theory
and read as $R_j=2 T_{0j}$. Finally, the $\{ij\}$ equations are:
\beq
G_{ij} + {1\over 2} m^2 (h_{ij}-a\delta_{ij}h) = T_{ij}\,.
\label{massij}
\eeq
We see that in the linearized theory the $\{00\}$ equation (\ref {mass00}) 
is also identical to the linearized massless Einstein equation.
Therefore, it is only the $\{ij\}$ equation that 
differentiates  (\ref {halfmass}) from the Einstein gravity in the 
linearized approximation.  Because of this one should expect
the vDVZ discontinuity to be absent. This can be checked in a rigorous
way. Let us follow the decomposition of Ref. \cite {Rubakov}:
\beq
h_{00}&=&\psi\,,\\
h_{0i}&=&u_i+\partial_i v \,,\\
h_{ij}&=&\chi_{ij}+\partial_{(i}s_{j)}+
\partial_i\partial_j\sigma+\delta_{ij}\tau
\label{decomp}\,,
\eeq
where $\chi_{ij}$ is a transverse-traceless tensor, $s_{j}$ is a 
transverse vector while the other fields are scalars.
The  gauge invariant combinations are:  a tensor $\chi_{ij}$,  
a vector $w_i=u_i-\partial_0 s_i$,  and two scalars $\tau$ and 
$\Phi=\psi-2\partial_0v+\partial_0^2\sigma$.  
The conventional coupling to a conserved matter stress-tensor
$h_{\mu\nu} T^{\mu\nu}$ can be written in terms of these invariants:
\beq
\chi_{ij}T_{ij} - 2 w_j T_{0j} + \Phi T_{00}+ \tau T_{jj}\,.
\label{couplings}
\eeq
Solving the corresponding linearized equations for $a\neq 1$ we find:
\beq
\chi_{ij}\,&=&\,{1\over - \partial_0^2+\Delta  -m^2}\,T^{tt}_{ij}\,,
\label{chi} \\
\tau \,&=&\, {1\over 2\Delta}T_{00}\,,~~~
w_j\,=\,{1\over \Delta}T_{0j}\,, \label{w} \\
\Phi \,&=&\,{1\over 2\Delta} \left (T_{jj}\,+\,T_{00}\, -\, 
{3\over \Delta}\partial_0^2\,T_{00}\right ) + {1-3a\over 2(1-a) \Delta^2}
\,m^2\,T_{00}\,.
\label{phi}
\eeq
(Here $ T^{tt}_{ij} $ stands for a transverse-traceless part of 
the tensor). These expressions give exactly the fields of the Einstein theory
in the limit $m\to 0$. Therefore, there is no vDVZ discontinuity.
Note that the only propagating degrees of freedom
are two polarizations of the transverse-traceless tensor 
$\chi_{ij}$. The spectrum is free of ghost and tachyons. 
Similar properties have been found previously 
in a stimulating work \cite {Dubovsky1}, where a somewhat 
different model was discussed\footnote{However, non-linear 
stability of the model of Ref. \cite {Dubovsky1}, which is a suspect
in the context of the discussions of our section 3, 
has not been studied in Ref. \cite {Dubovsky1}.}.

On the other hand, the above system exhibits the same type 
of instantaneous interactions as  QED' discussed in Section 2. 
This is because the instantaneous parts cannot be exactly canceled
as long as there is a mass term in the denominator of (\ref {chi}).
However, as we discussed  in the gauge field case, 
these instantaneous interactions can only be probed by  
very low-momentum/frequency sources. It would be interesting to study 
the phenomenology of this model \cite {GLuca}. \\

{\it 2. More degrees of freedom} \\
A generalization of the above model can be obtained
by adding to (\ref {halfmass}) a mass term for $N_j$:
\beq
\Delta \l_2= {c} \,m^2\,  \sqrt{\gamma}N_i^2\,,
\label{halfmass1}
\eeq
where $c$ is a constant. Doing so we add three additional degrees of 
freedom to the theory. The Hamiltonian now takes the form:
\beq
\mathcal{H}=\sqrt{\gamma}\left[
	N\left(R^0+\frac{1}{2}m^2(h_{ij}^2-ah^2)\right)+
	N_i\left(R^i-  m^2c\,N^i\right)\right]\,.
\label{ham3}
\eeq
The corresponding constraint equations are:
\beq
R^0& =& -\frac{1}{2}m^2(h_{ij}^2-ah^2)\,, \label{con1} \\
R^i&=&2m^2c\,N^i\,.  \label{con2}
\eeq
As long as $c\neq 0$ 
the only true constraint  is (\ref{con1}), since (\ref{con2}) 
does not restrict  the number of propagating degrees of freedom but
acts as an algebraic equation determining $N^j$. 
Therefore, this counting tells us that the number of propagating degrees
of freedom is five.  Solving (\ref{con2}) for $N^i$ and substituting 
the result in  $\mathcal{H}$ we find the Hamiltonian 
\beq
\label{H}
\mathcal{H}=\sqrt{\gamma} \frac{R_i^2}{4\,m^2\,c}\,, 
\label{hamc}
\eeq
which is positive semidefinite as long as $c>0$. This model
also has no vDVZ discontinuity. The calculations are 
similar to those presented above but more tedious, 
the spectrum contains no ghost of tachyons.
The massless limit of (\ref {ham3}) is regular and one 
recovers in this limit the Einstein gravity. Moreover,
the theory is symmetric w.r.t. spatially independent 
transformations of the time variable, and, exhibits 
the instantaneous interactions.Further interesting 
properties of this model will be discussed in \cite {GLuca}. \\

One can also  evade the BD instability and the presence of 
the long range interactions by adding into the Lagrangian 
yet another  term
\beq
\Delta\l_3\,=\, m^2\sqrt{\gamma}\,P_2(N)\,,
\label{nolongrange}
\eeq
where $P_2(N)$ is a polynomial in $N$ of degree 2, 
namely $P_2(N)=c_0+c_1N+c_2N^2$. We require that  there are no 
constant and linear 
terms in the linearized Lagrangian. This gives  the relations 
$c_0+c_1+c_2 =0$ and $c_2+\frac{1}{2}c_1=0$, with a solution 
$c_1=-2c_0$ and $c_2=c_0$. Hence, $P_2(N)=c_0(N-1)^2$.
The Hamiltonians of the system is that of the previous examples
plus the new term $\Delta\mathcal{H}=-m^2\sqrt{\gamma}P_2(N)$. 
The resulting constraint equations are:
\beq
R^0+\frac{1}{2}m^2(h_{ij}^2-ah^2)&=&m^2P_2^\prime(N)\,,\\
R^i&=&2 m^2c\,N^i\,.
\eeq
Solving the constraints w.r.t. $N$ and $N^i$ we obtain
\beq
N&=& \frac{R^0+\frac{1}{2}m^2(h_{ij}^2-ah^2)+2c_0m^2}{2c_0m^2}\,,\\
N^i&=&\frac{R^i}{2m^2c}\,,
\eeq
and the Hamiltonian
\begin{equation}
\mathcal{H}=\sqrt{\gamma}\left[R^0_\mathrm{mod}+
\frac{({R^0_\mathrm{mod}})^2}{4c_0m^2}+\frac{R_i^2}{4m^2c}\right]\,,
\end{equation}
where $R^0_\mathrm{mod}\equiv R^0+\frac{1}{2}m^2(h_{ij}^2-ah^2)$.
The above Hamiltonian is not necessarily positive semidefinite, however,
it is bounded from below as long as $c,c_0>0$. 
There are six degrees of freedom propagating in this model. 
This could exclude the model based on the Solar system data.
However, it is not  impossible to imagine that the nonlinear effects
suppress the couplings of the extra polarizations to matter 
at observable distances, in analogy with \cite {DDGV,Iglesias}.
A more serious  problem of the model 
(\ref {nolongrange}) is the absence of a symmetry principle that 
would guarantee the stability w.r.t. quantum corrections.

\vspace{0.3in}

\section{Acknowledgments}

We would like to thank Gia Dvali, Massimo Porrati, Matthew Schwartz
and  Dan Zwanziger for useful discussions. 
The work of GG is partially supported 
by the NSF. LG is supported by  Physics Department Graduate 
Student Fellowship at NYU.

\end{document}